\documentclass[aps,preprint,prapplied,superscriptaddress]{revtex4}
\usepackage{graphicx}
\usepackage{dcolumn}
\usepackage{bm}
\usepackage[usenames]{color}
\usepackage{amsmath}
\usepackage{amssymb}
\usepackage{latexsym}

\begin{document}

\title{Emergence of the Chern Structure Using Sr$_2$RuO$_4$ Nanofilms}

\author{Hiroyoshi Nobukane}
\affiliation{Department of Physics, Hokkaido University, Sapporo, 060-0810, Japan}
\affiliation{Center of Education and Research for Topological Science and Technology, Hokkaido University, Sapporo, 060-8628, Japan}

\author{Toyoki Matsuyama}
\affiliation{Center of Education and Research for Topological Science and Technology, Hokkaido University, Sapporo, 060-8628, Japan}
\affiliation{Department of Physics, Nara University of Education,
Nara 630-8528, Japan}

\author{Satoshi Tanda}
\affiliation{Center of Education and Research for Topological Science and Technology, Hokkaido University, Sapporo, 060-8628, Japan}

\affiliation{Department of Applied Physics, Hokkaido University, Sapporo 060-8628, Japan}

\date{\today}

\begin{abstract}
We discovered a fractional Chern structure in chiral superconducting Sr$_2$RuO$_4$ nanofilms by employing electric transport.
By using Sr$_2$RuO$_4$ single crystals with nanoscale thickness, a fractional Hall conductance was observed without an external magnetic field.
The Sr$_2$RuO$_4$ nanofilms enhanced the superconducting transition temperature to about 3~K.
We found an anomalous induced voltage with temperature and thickness dependence, and the switching behavior of the induced voltage appeared when we applied a magnetic field. 
We suggest that there was fractional magnetic-field-induced electric polarization in the interlayer.
These anomalous results are related to topological invariance.
The fractional axion angle $\theta=\pi/6$ is determined by observing the topological magneto-electric effect in Sr$_2$RuO$_4$ nanofilms.

\end{abstract}

\maketitle
\section{Introduction}

The mathematical structure characterized by Chern numbers~\cite{Chern} has yielded very important findings in both condensed-matter and high-energy physics~\cite{Volovik}.
The quantum Hall effect in graphene provides the quantized Hall conductance of $\sigma_{H}=(p/q)e^2/h$ with $p$ and $q$ coprimes~\cite{Novoselov}. 
The quantization of the Hall conductance and anyonic particles with exotic mutual statistics can be explained from the Chern-Simons (CS) term in (2+1)-dimensional topological field theory~\cite{Semenoff,Matsuyama,Haldane}. 
Recently, the topological magneto-electric effect in topological insulators and superconductors has been predicted theoretically from the Chern-Pontryagin term (topological $\theta$-term with $\theta=\pm\pi$) in 3+1 dimensions~\cite{Qi_topo,Qi_axion}.
In strongly correlated superconductors and insulators, the presence of fractional Chern structures is particularly expected based on an analogy with the relation between the integer and fractional quantum Hall effects~\cite{Swingle}.
The Chern structures may clarify the universality of topology in nature.
However, as yet there has been no experimental evidence of Chern structures in superconductors.
The Chern structure in superconductivity is interesting in itself and may also have implications regarding the esoteric physics of quantum chromodynamics (axion electrodynamics)~\cite{Witten,Wilczek} in  condensed-matter experiments.
In addition to the purely scientific interest that they arouse, these quantum phenomena may provide the basis for new applications to magneto-electric coupling devices and for the development of a topological quantum computation of non-Abelian statistics~\cite{Nayak}.

Layered perovskite Sr$_2$RuO$_4$ is a leading candidate for a spin-triplet and chiral $p$-wave superconductor in quasi-two-dimensional electron systems~\cite{RMP_SRO}, and is also known as a Chern superconductor with a non-zero Chern invariant. 
The spontaneously broken time-reversal and parity symmetry realize novel topological quantum phenomena such as zero-magnetic-field quantum Hall effects~\cite{goryo,Volovik_JETP,volovik_yako}, gapless Majorana excitations in an edge or the core of vortices~\cite{Read} and the non-Abelian statistics of half-quantum vortices~\cite{Ivanov}.
However, the chiral-multi-domains in millimeter-scale Sr$_2$RuO$_4$ obscure these novel phenomena.
In the microscale chiral single domain of Sr$_2$RuO$_4$ with submicron thickness, we have reported that the current-voltage ($I-V$) curves violate parity due to the excitation of the Majorana-Wyle fermions along the one-dimensional chiral edge current~\cite{nobukane_PRB,nobukane_SSC}.
To clarify the Chern structure through quantum transport in units of $e^2/h$ in two-(or quasi-two-) dimensional chiral superconducting layers, we have investigated electric transport properties by using chiral single domain size Sr$_2$RuO$_4$ with nanoscale thickness~\cite{nobukane_physica-B}.
Specifically, in this paper, we report the emergence of the fractionalized Chern structure (number) from the results of the anomalous properties, which are revealed by reducing the Sr$_2$RuO$_4$ thickness to the nanometer range.

In Sr$_2$RuO$_4$ systems, one unsolved problem is the two superconducting phases with $T_c\sim 1.5$ and 3~K.
Although the pure Sr$_2$RuO$_4$ single crystals exhibit a $T_{c}$ of about 1.5~K, enhancement to about 3~K has been reported in Sr$_2$RuO$_4$-Ru eutectic systems~\cite{maeno_JPSJ}. 
However, recent investigations have found that, even in pure Sr$_2$RuO$_4$ without Ru inclusions, an enhanced $T_c$ of around 3~K was observed when measuring uniaxial pressure effects along the $c$ axis~\cite{Kittaka}, strain effects~\cite{Hicks} and the properties near the lattice dislocations~\cite{Liu}. 
As regards this discrepancy, electric transport measurements in nanoscale thin films of Sr$_2$RuO$_4$ single crystals allow access to both topological quantum states and the pairing mechanism itself in chiral $p$-wave superconductors.

In this paper, we report the emergence of a fractionalized Chern structure in Sr$_2$RuO$_4$ single crystal nanofilms based on the anomalous transport properties observed for the in-plane and interlayer directions.
By reducing the Sr$_2$RuO$_4$ thickness to the nanometer range, we found that a fractional quantum Hall resistance of $h/4e^{2} - h/2e^{2}$ occurred as a consequence of the spontaneous Hall current without an external magnetic field in the chiral $p$-wave superconductivity.
The nanofilms exhibited an enhanced $T_c$ of about 3~K.
The anomalous induced voltage and the switching behavior were observed as a function of temperature and thickness under zero bias current for an applied magnetic field parallel to the $c$ axis.
The applied magnetic field induced electric polarization in the superconducting interlayer of Sr$_2$RuO$_4$ nanofilms.
In Sr$_2$RuO$_4$ with a nanoscale thickness, we demonstrated the fractional topological magneto-electric effect in three (quasi-two) dimensions, which is characterized by the fractional axion angle (coefficient) $\theta=\pi/6$ of the {\boldmath$E$}$ \cdot ${\boldmath$B$} term.

\section{Experiment}

To obtain nanoscale Sr$_2$RuO$_4$ thin films, we synthesized Sr$_2$RuO$_4$ single crystals with a solid phase reaction, and selected single crystals with no embedded Ru metal by observing optical microscope images, chemical composition and  crystal orientation~\cite{nobukane_jjap}. 
Sr$_2$RuO$_4$ single crystal nanofilms were exfoliated on a SiO$_2$(300~nm)/Si substrate. 
We then fabricated gold electrodes using standard electron beam lithography methods.
Scanning electron micrographs of the Sr$_2$RuO$_4$ nanofilms (samples A and B) are shown in Fig.~\ref{figure1}(a) and \ref{figure1}(b).
Samples A and B were 17~nm and 20~nm thick, respectively.
For several samples with thicknesses of $17-400$~nm, electric transport properties were measured by the four-terminal method using a homemade $^3$He refrigerator. 
The longitudinal and Hall voltages and the differential conductance were measured with a nanovoltmeter (2182, Keithley) and a lock-in-amplifier (5210, Princeton Applied Research), respectively.
The up and down magnetic field sweep rate was 0.102 mT/sec.

\section{Results and Discussion}

The measurements of the nanoscale Sr$_2$RuO$_4$ thin films revealed a fractional Hall resistance. 
Figure~\ref{figure1}(a) shows the temperature dependence of the Hall resistance $R_{xy}$ and longitudinal resistance $R_{xx}$ for bias current $I=\pm$10~nA.
In general, bulk (thick) superconductors exhibit zero resistance below $T_c$.
We have observed zero resistivity below $T_c=1.59$~K for microscale Sr$_2$RuO$_4$ with a thickness of 340~nm~\cite{nobukane_PRB}, which is consistent with the $T_c$ of bulk Sr$_2$RuO$_4$ crystals~\cite{RMP_SRO, maeno_JPSJ}.
On the other hand, two-dimensional superconductors allow a quantum resistance of $h/4e^2$~\cite{Haviland, Tanda, Bollinger}.
In sample A, with decreasing temperature, the Hall resistance increased with a log$T$ dependence, and reached about 12.1~k$\Omega$ below 0.8~K as shown in Fig.~\ref{figure1}(a). 
Interestingly, in the absence of an external magnetic field, we measured a Hall resistance of 12.1~k$\Omega$, which is close to the quantum resistance of $h/2e^2$.
A Hall resistance of $R_{xy}=6.8$~k$\Omega\sim h/4e^2$ was also observed in sample B as shown in Fig.~\ref{figure1}(b).
The quantum Hall resistance was reproduced.
We note that the $R_{xx}$ values of samples A and B were 5.3~k$\Omega \sim h/5e^2$ and $6.1$~k$\Omega \sim h/4e^2$ at lower temperature.
These are very close to the quantum critical sheet resistance of 6.45~k$\Omega=h/(2e)^2$ in a two-dimensional superconductor-insulator transition~\cite{Fisher}.
Thus we think that the quantum resistance in Sr$_2$RuO$_4$ nanofilms is related to the dynamics of Cooper pairs and vortices in two dimensions.
The $R_{xy}$ and $R_{xx}$ values in the thick samples with thicknesses of 400 and 340~nm exhibited much smaller values of $ 0.1 \sim 1~\Omega$ than the quantum resistance in the thin film samples.

By varying the thickness of the film, we obtain different $T_c$ values of 1.5 and 3~K.
Figure~\ref{figure1}(c) shows the $I-V_{xy}$ characteristics of the Hall bar geometry in sample A, and $dI/dV_{xy}$ as a function of the Hall voltage $V_{xy}$ in zero magnetic field at several temperatures, which is vertically shifted for clarity.
The Hall conductance spectra at 0.43~K exhibited a superconducting gap $\Delta$ with coherence peaks at $\sim \pm 0.4~$meV.
This is comparable to the superconducting gap size in previous reports on tunneling spectroscopy in Sr$_2$RuO$_4$ systems~\cite{Laube, Mao, Suderow, Kashiwaya, Davis}.
The temperature dependence of the superconducting gap extracted from the coherence peak width in the tunneling spectra is shown in the inset of Fig.~\ref{figure1}(c).
Below 3~K, clear dip structures were observed.
Here a 400-nm-thick Sr$_2$RuO$_4$ single crystal (sample C) shows neither the suppression of $T_c$ nor an enhancement to 3~K as shown in Fig.~\ref{figure1}(d).
The $T_c$ for sample C is evidence of good quality single crystals.
The conductance peak in the vicinity of zero bias current in the superconducting state may prove the existence of a gapless chiral Majorana state.
The position of the conductance peak was shifted to a minus bias current below $T_c$.
In recent studies, pure Sr$_2$RuO$_4$ systems have enhanced the $T_c$ to 3~K~\cite{maeno_JPSJ,Kittaka,Hicks,Liu}.
We believe that Sr$_2$RuO$_4$ nanofilms provide the nucleation needed for surface superconductivity below 3~K.
Thus, a superconducting gap below 3~K appears to be an intrinsic property of Sr$_2$RuO$_4$.
As a consequence of this significant feature of Sr$_2$RuO$_4$ superconductivity, we observed zero-magnetic-field anomalous Hall conductance.

To examine the topological magneto-electric coupling in Sr$_2$RuO$_4$ nanofilms, we measured the voltage $V$ at zero bias current.
Figure~\ref{figure2}(a) shows the magnetic field dependence of the voltage when a magnetic field is applied parallel to the $c$ axis from zero magnetic field up to $\pm$7~T for sample A.
Interestingly, we found an anomalous induced voltage of about $|\Delta V|=62~\mu$V in zero magnetic field at 0.43~K.
In the superconducting state below 3~K, the anomalous voltage appeared as shown in the inset of Fig.~\ref{figure2}(b), which is consistent with the $V-B$ characteristics in Fig.~\ref{figure2}(a).
Figure~\ref{figure2}(b) shows the thickness dependence of the induced voltage at lower temperature.
As the sample thickness $t$ was reduced, the induced voltage increased, which is fitted well by the relation $V=1/t$.
Since spontaneous voltage is related to sample thickness, we can eliminate the contribution of the thermoelectric voltage and a junction at the Au/Sr$_2$RuO$_4$ interface or at a microcrack.
Furthermore, the switching behavior of an induced voltage was observed in the region between $\pm$1.0 and $\pm$5.8~T.
The solid red line in Fig.~\ref{figure2}(a) is the averaged result for the $B-V$ characteristics.
Figure \ref{figure3}(a) shows the magnetic field dependence of the induced voltage and the switching voltage at various temperatures.
With increasing temperature, the anomalous induced voltage and the switching voltage were gradually suppressed, and vanished above 3~K.
In sample C with a thickness of 400~nm, we also observed an induced voltage of 1~$\mu$V and switching behavior under an applied magnetic field as shown in Fig.~\ref{figure2}(d).
The anomalies became smaller than the results observed for nanofilms (sample A).
Table~\ref{table1} summarizes the properties obtained by varying the thickness.

\begin{table}[h]
\begin{tabular}{lcc}
\hline\hline
 & Sample A & Sample C \\

\hline

Thickness (nm) & 17 & 400 \\

$T_c$ (K) & 3 & 1.5 \\

$R_{xy}$ (k$\Omega$) & $12$ & 1~$\times~10^{-3}$ \\

Induced $V$ ($\mu$V) & 62 & 1\\

$B$ at maximal $V_{SW}$ (T) & 3.5 & 0.02 \\

$\Delta E \cdot \Delta B$ ($h/e^2$) & 6 & 0.5~$\times~ 10^{-3}$ \\

\hline\hline
\end{tabular}

\caption{Summary of the properties of Sr$_2$RuO$_4$}
\label{table1}
\end{table}


To analyze the switching phenomena in more detail, we subtract the averaged $B-V$ curves from the measured $B-V$ characteristics.
The results for sample A are shown in Figs.~\ref{figure2}(c) and~\ref{figure3}(b).
Here the voltage $V_{SW}$ represents the switching voltage component of the induced voltage.
With increasing magnetic field, the anomalous switching voltage $V_{SW}$ increases above $\sim \pm$1~T.
The $V_{SW}$ for the applied magnetic field reaches a maximum value near $\pm$3.5~T, and then decreases.
Intriguingly, the switching voltage was clearly observed below 1.5~K. 
We think that the observation of the anomalous switching voltage is related to the intrinsic properties of the chiral $p$-wave superconductor Sr$_2$RuO$_4$, because the feature appears below a $T_c$ of about 1.5~K in bulk Sr$_2$RuO$_4$.

We discuss the enhancement of the critical magnetic field in relation to the surface superconductivity of Sr$_2$RuO$_4$.
In sample C, the magnetic field of 0.04~T caused the anomalous $V$ and $V_{SW}$ to vanish, which is consistent with $\mu_{0}H_{c2}$ in bulk Sr$_2$RuO$_4$.
On the other hand, in Figs.~\ref{figure2}(a) and~\ref{figure3}(a), the anomalous voltage is induced even in a magnetic field beyond $\mu_{0}H_{c2}$ reported for pure bulk Sr$_2$RuO$_4$. 
To reveal whether or not the anomalous behavior is an intrinsic characteristics of Sr$_2$RuO$_4$, we need to consider the physical properties of the 3~K phase.
The 3~K superconductivity in Sr$_2$RuO$_4$-Ru systems induces the enhancement of the upper critical field to $\mu_{0}H_{c2//c}(0) \approx 1.5$~T and $\mu_{0}H_{c2//ab}(0) \approx 4$~T~\cite{RMP_SRO}, respectively.
This may be comparable to the results for the $B-V$ characteristics in our nanofilm samples.
In addition to being the origin of the enhancement of $T_{c}$ up to 3~K, the surface superconductivity effect also enhances the upper critical field in pure Sr$_2$RuO$_4$ nanofilms.
Thus, we believe that anomalous behaviors such as the Hall conductance, the induced voltage and the switching, can be explained by considering the topological magneto-electric effect as discussed below.

To understand the origin of the above behaviors, we address (I) the fractional quantum Hall conductance in the conducting layer without a magnetic field, (I$\hspace{-.1em}$I) the fractional magnetic-induced electric polarization in the interlayer, and (I$\hspace{-.1em}$I$\hspace{-.1em}$I) the integral value of $E \cdot B$ in the topological term.  
First, let us discuss the fractional quantum Hall conductance in zero magnetic field in Sr$_2$RuO$_4$ nanofilms.
From $R_{xy}=$ 12.1 k$\Omega$ and $R_{xx}=5.3$ k$\Omega$ for sample A, the sheet Hall conductance $\sigma^{n_s=1}_{xy}=(1/22.6)~e^2/h\sim(1/24)~e^2/h$ per RuO$_2$ layer was determined by using the relation $R^{n_{s}=1}_{xy}=\rho_{xy}^*/d$, $\rho_{xy}^*=(\sigma_{xy})^{-1}=\left({\rho_{xy}}/{(\rho_{xx}^2+\rho_{xy}^2)}\right)^{-1}$, $d=6~$\r{A}~$(=c/2)$.
By considering the number of sheets $n_{s}$, we represent the Hall conductance as $\sigma^{n_{s}}_{xy}=\sigma^{n_s=1}_{xy}n_{s}=(n_{s}/24) e^2/h=(R^{n_{s}}_{xy})^{-1}$.
Why do Sr$_2$RuO$_4$ nanofilms exhibit Hall conductance quantized in the unit of conductance quantum $G_{0}\equiv e^2/h$?
The concept of chiral $p$-wave superconductivity can be developed by the induced CS-term of the effective Lagrangian in topological field theory~\cite{goryo,Volovik_JETP}.
The CS-term induces the existence of a spontaneous Hall current in a zero magnetic field perpendicular to the bias current direction.
However, the quantum Hall effect in Sr$_2$RuO$_4$ has yet to be observed experimentally for the following reasons.
The Hall resistance in a thick sample becomes smaller than that in a thin film sample, and the observation of the ensemble averaging of the Hall current in multi-chiral domains is complicated.
With respect to these issues, an important solution is for the sample to consist of nanoscale thin films of~$\sim$~10~layers~\cite{Volovik,volovik_yako} and for the chiral single domain size to be $\sim 1~\mu$m~\cite{nobukane_PRB}.
Our samples satisfy these conditions.
Thus, by using Sr$_2$RuO$_4$ nanofilms, we can observe the fractional quantum Hall conductance in chiral $p$-wave superconductivity.

From the result of the fitted slope of $|\Delta V|/\Delta B$ in Fig.~\ref{figure2}(a), we discuss the contribution of the electric polarization under a magnetic field in the interlayer of Sr$_2$RuO$_4$.
We assume that our layered sample is a superconducting bilayer system in order to discuss the possibility of magneto-electric polarization.
The capacitance $C=\varepsilon_{I}A/d=10$ fF of the interlayer is estimated, where $\varepsilon_{I}/\varepsilon_{0} \sim 10$ is the interlayer dielectric constant, $d~(=6~$\r{A}) is the interlayer distance in Sr$_2$RuO$_2$, and $A$ is the area 0.14~$\mu$m$^2$ between the electrodes.
We determined the effective electric charge $Q^* \sim 8~e$ from the induced voltage of $|\Delta V|=$ 63 $\mu$V.
Moreover, we obtained the effective magnetic flux $\Phi^* \sim 206~\Phi_0$ from the relation $|\Delta V|/\Delta B = Q^{*}d/\varepsilon_{I}\Phi^{*}$, where $\Phi_{0}=h/2e$ is the magnetic flux quantum. 
We found the fractional magneto-electric polarization {\boldmath$P$}/{\boldmath$B$}$ = (1/12)e^2/h~(-(1/12)e^2/h)$ from the slope in the positive (negative) magnetic field.
Surprisingly, the fractional coefficient of the magnetic-field-induced-electric polarization is equivalent to that of the Hall conductance in the bilayer film.

Below, we consider the topological magneto-electric effect in Sr$_2$RuO$_4$ nanofilms to understand the relationship between fractional Hall conductance and electric polarization.
The chiral anomaly~\cite{Adler, Bell-Jackiw} in the (3+1)-dimensional topological field theory can introduce an additional $\theta$-term
$S_\theta=\frac{\theta}{2\pi}\frac{e^2}{h}\int d^{3}x dt {\mbox{\boldmath $E$}} \cdot {\mbox{\boldmath $B$}}$~\cite{Qi_topo, Qi_axion}.
This topological term denotes the existence of the magneto-electric effects.
Namely, an applied electric field generates a magnetic polarization {\mbox{\boldmath $M$}}$=\frac{\theta}{2\pi} \frac{e^2}{h}${\mbox{\boldmath $E$}, and an applied magnetic field generates an electric polarization  {\mbox{\boldmath $P$}}$=\frac{\theta}{2\pi} \frac{e^2}{h}${\mbox{\boldmath $B$}.
The quantum Hall current {\mbox{\boldmath $J_{H}$}}$=\frac{\theta}{2\pi} \frac{e^2}{h}${\mbox{\boldmath $E$} flows as the contribution of the topological surface state in the (3+1)-dimensional magneto-electric effect.
For strongly correlated electron systems, the fractional parameter $\theta=p/q$ with $p, q$ odd integers is predicted by analogy with fractional quantum Hall effects~\cite{Swingle,Maciejko}.
This model of fractionalization in the chiral anomaly appears to be beneficial in terms of understanding our anomalous results.

Furthermore, we discuss the possibility of the fractionalization of the topological $\theta$-parameter.
Using the experimental results, we discuss the integral value $\int d^{3}x dt~{\mbox{\boldmath $E$}} \cdot {\mbox{\boldmath $B$}}$ in the topological $\theta$-term.
We found that the value of $E{(=\Delta V/d)} \cdot B$ represented by the blue square region $A_\Box$ in Fig.~\ref{figure3}(a) is equivalent to that of $E \cdot B$ represented by the red rhombic region $A_\Diamond$ in Fig.~\ref{figure3}(a) and (b).
An important point is that the obtained value of $E\cdot B$ is $6(h/e^2$) at 0.43~K.
Similarly, at temperatures below 2.0~K, we estimated the value of $E\cdot B$ in both the square region in a low magnetic field  and the rhombic region provided by the voltage switching. 
We confirmed that the $E\cdot B$ values are the same in the blue and red regions at each temperature.
The correspondence of the $E\cdot B$ value is also reproduced in sample C in Fig~\ref{figure2}(d).
This means that the induced voltage in a low magnetic field is closely related to the occurrence of the switching voltage under a magnetic field, and these are connected to the topological invariant.
Figure~\ref{figure3}(c) shows the temperature dependence of the $E\cdot B$ value.
The data points were fitted by an exponential curve.
We believe that the $E\cdot B$ value exists in the $6(h/e^2)-12(h/e^2)$ region. 
According to the fitting curve, at $T=0$, (I$\hspace{-.1em}$I$\hspace{-.1em}$I) the $E\cdot B$ value is about $12h/e^2$, which is comparable to (I) the zero-magnetic-field quantum Hall conductance $\sigma^{n_{s}=2}_{xy}=(1/12)e^2/h$ and (I$\hspace{-.1em}$I) the magnetic-field-induced electro-polarization {\boldmath$P$}/{\boldmath$B$}$=(1/12)e^2/h$.
By substituting $\int d^{3}x dt E\cdot B = 12h/e^2$ into the topological term $S_\theta$, we obtained the fractional angle $\theta=\frac{\pi}{6}N$, where $N$ is an integer multiple.
For the fractional angle  $\theta=\pi/6$, the quantum Hall conductance $\sigma_{xy}=\frac{\theta}{2\pi}\frac{e^2}{h}=\frac{1}{6}\frac{e^2}{2h}$, and electric polarization ${\mbox{\boldmath $P$}}=\frac{\theta}{2\pi}\frac{e^2}{h}{\mbox{\boldmath $B$}}=\frac{1}{6}\frac{e^2}{2h}{\mbox{\boldmath $B$}}$ are discussed theoretically.~\cite{Qi_topo, Maciejko}. 
This is consistent with our experimental observations.
Thus, we demonstrated the fractional topological magneto-electric effect in Sr$_2$RuO$_4$ layers using the results of both the fractional Hall conductance and the fractional magneto-electric polarization in the fractional axion angle.
Namely, these observations correspond to the fractional Chern structure in Sr$_2$RuO$_4$ superconductors.

\section{Summary}

In summary, we have detected the emergence of the fractional Chern structure in chiral superconducting Sr$_2$RuO$_4$ nanofilms by observing the fractional Hall conductance on the surface and the fractional electric polarization in the interlayer.
In a zero magnetic field, a quantized fractional Hall resistance was observed in the superconducting state below 3~K.
Under zero bias current, we found the anomalous induced voltage and the switching behavior of the induced voltage for an applied magnetic field parallel to the $c$ axis. 
The applied magnetic field generated electric polarization in the interlayer of Sr$_2$RuO$_4$.
The results suggest the presence of the fractional topological magneto-electric effect in Sr$_2$RuO$_4$ nanofilms.
The fractional axion angle $\theta=\pi/6$ in the topological $\theta$-term was also determined.

\begin{acknowledgements}

We thank Y. Asano, K. Inagaki, T. Honma, K. Ichimura, S. Takayanagi, K. Yamaya, N. Matsunaga, and K. Nomura for experimental help and useful discussions. 
This work was supported by JSPS KAKENHI (No. 25800183 and No. 26287069) and Takayanagi Memorial Foundation. 

\end{acknowledgements}

\newpage

\begin{figure}[t]
\begin{center}
\includegraphics[width=0.49\linewidth]{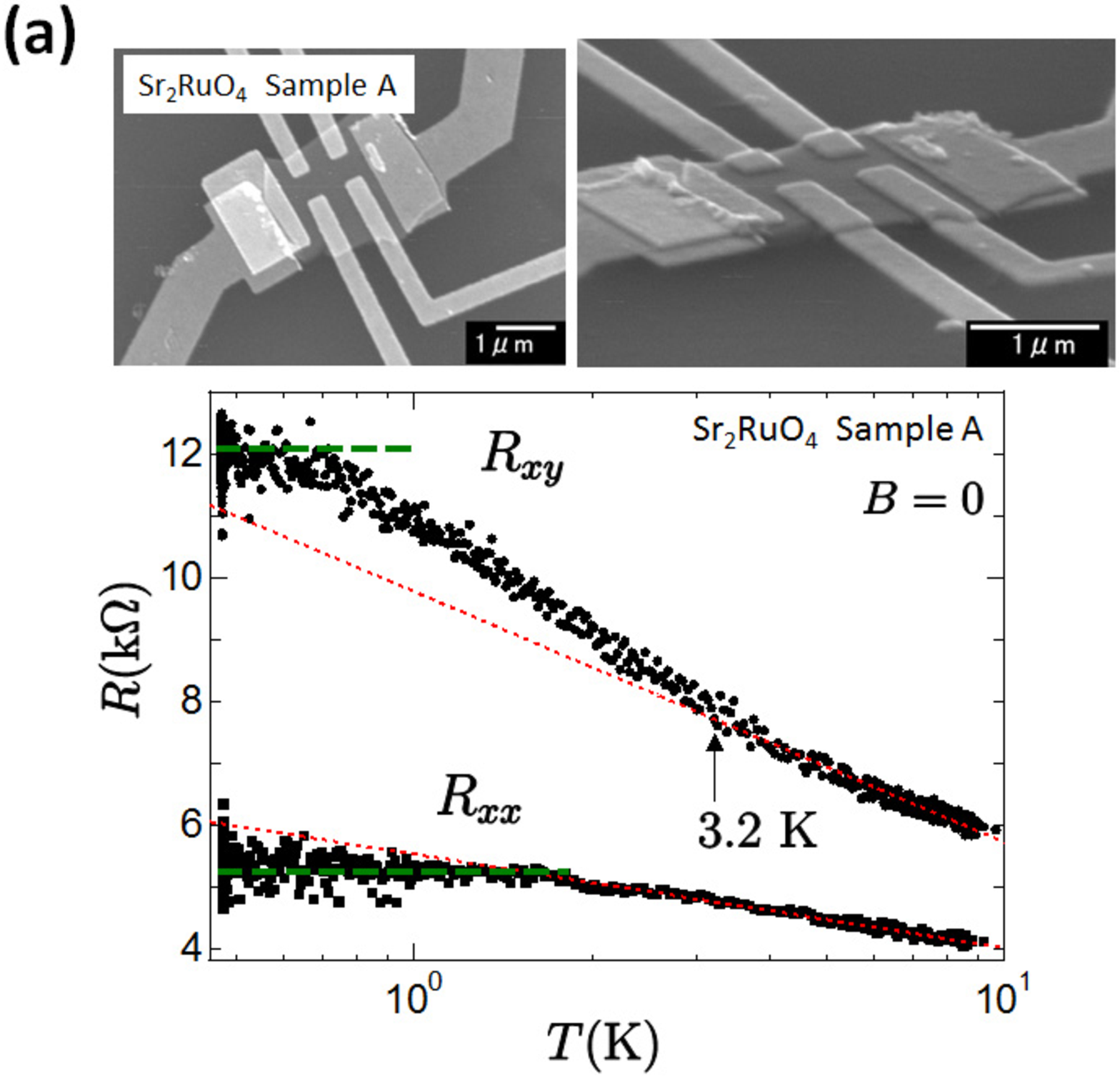}
\includegraphics[width=0.49\linewidth]{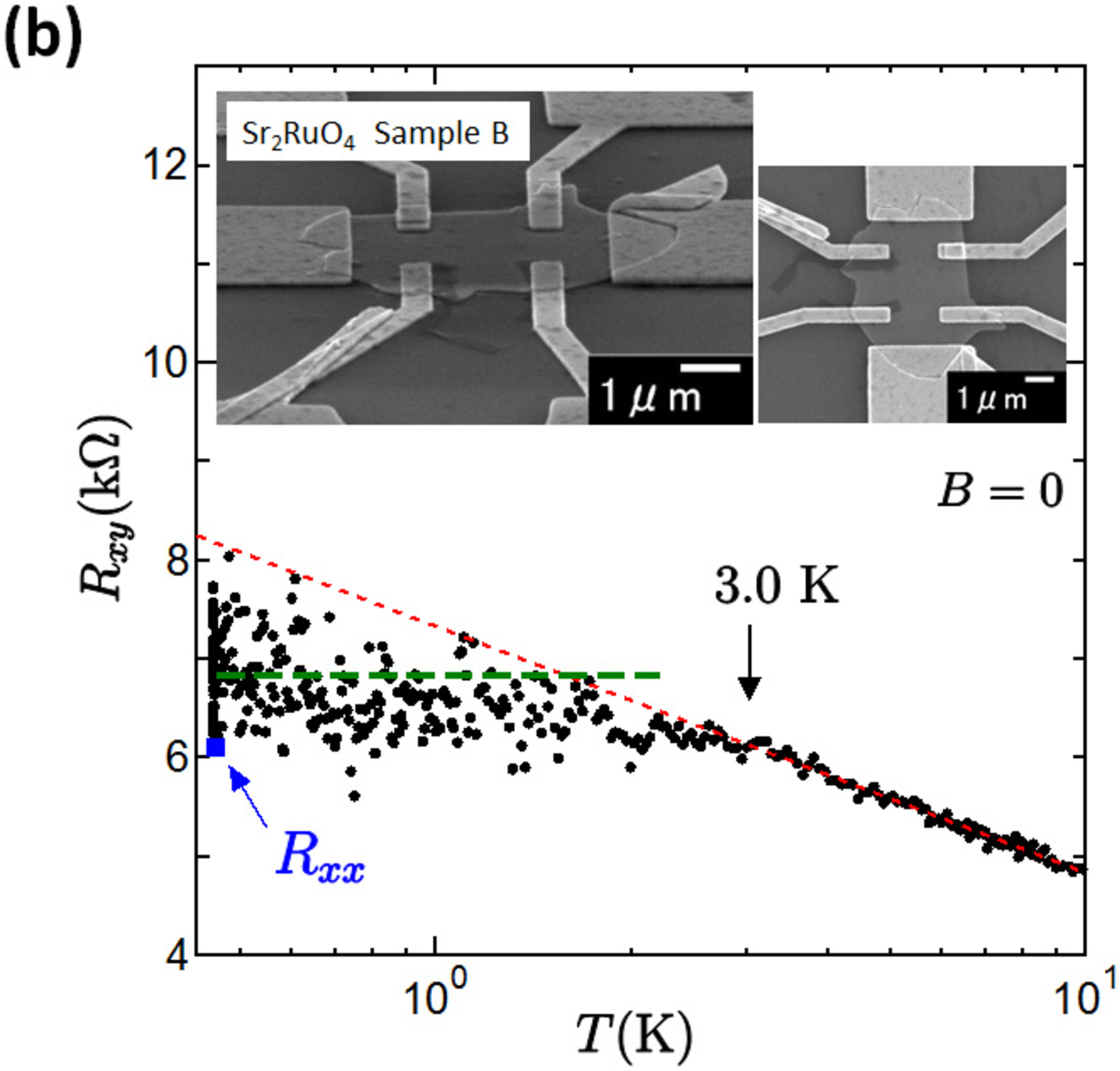}
\includegraphics[width=0.5\linewidth]{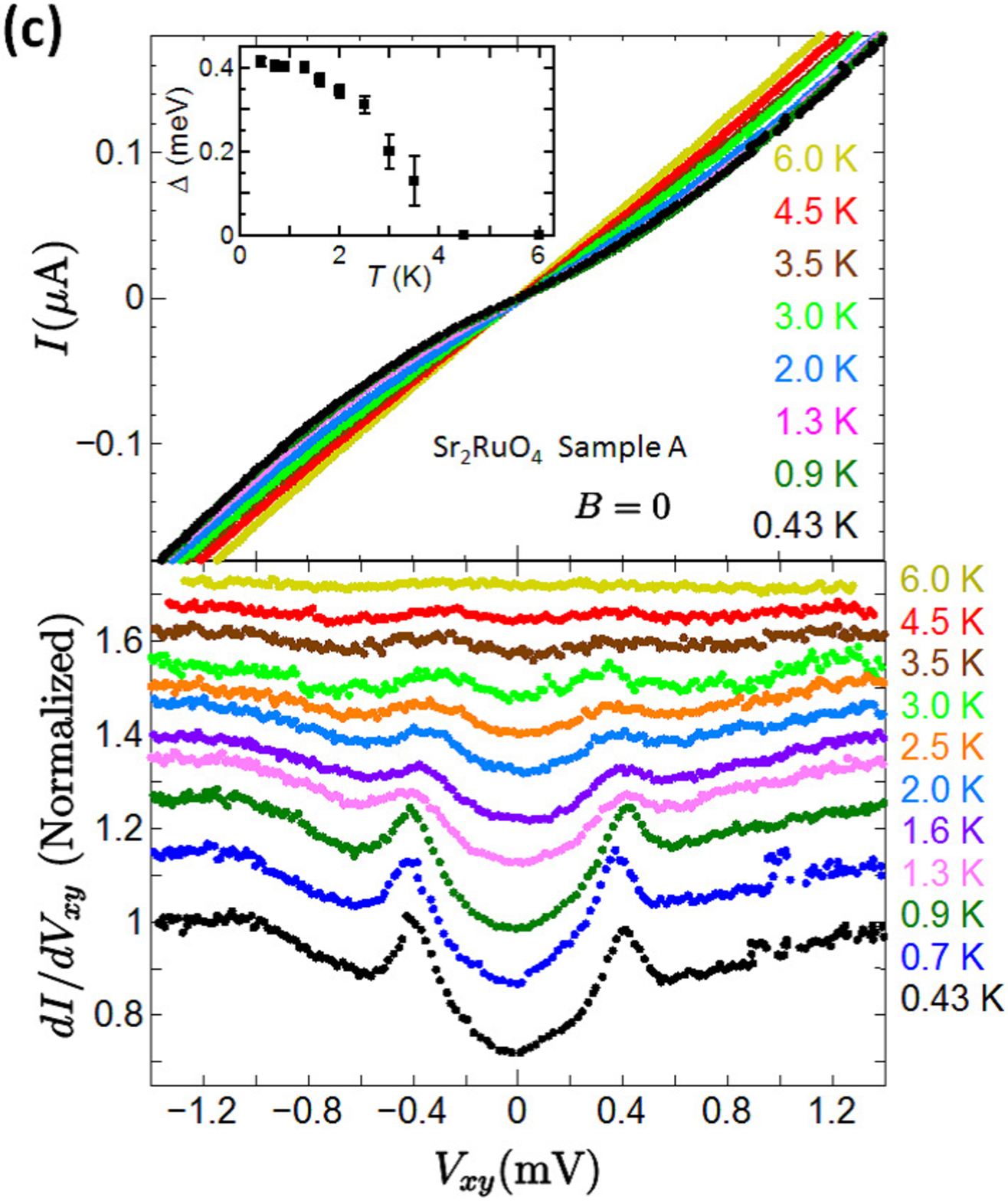}
\includegraphics[width=0.48\linewidth]{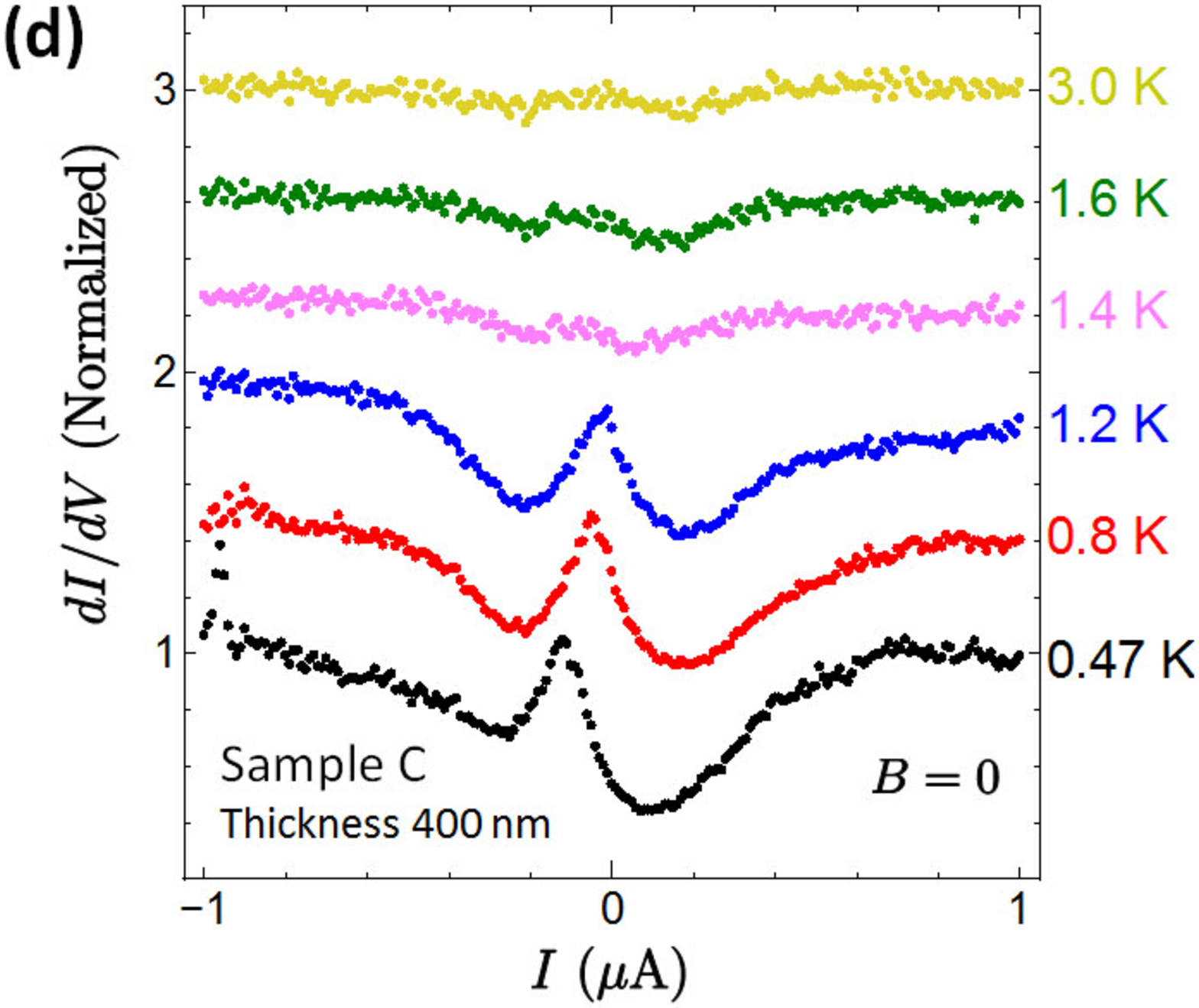}
\caption{
(a)Scanning electron micrographs of the top and side views of sample A.
Temperature dependence of $R_{xy}$ and $R_{xx}$.
The dotted horizontal lines are a guide.
(b)Temperature dependence of $R_{xy}$ for sample B. 
The solid square represents the result of $R_{xx}$ at 0.46~K.
The insets show micrographs of sample B. 
(c)$V_{xy}-I$ characteristics for sample A at temperatures in a zero magnetic field. 
$dI/dV_{xy}$ as a function of $V_{xy}$. 
The inset shows the temperature dependence of the superconducting gap $\Delta$.
(d)Differential conductance $dI/dV$ versus bias current for several temperatures in the Sr$_2$RuO$_4$ single crystal with a thickness of 400~nm (sample C), which is vertically shifted for clarity. The zero bias conductance peak was found below $T_c$.
}
\label{figure1}
\end{center}
\end{figure}

\newpage

\begin{figure}[t]
\begin{center}
\includegraphics[width=0.49\linewidth]{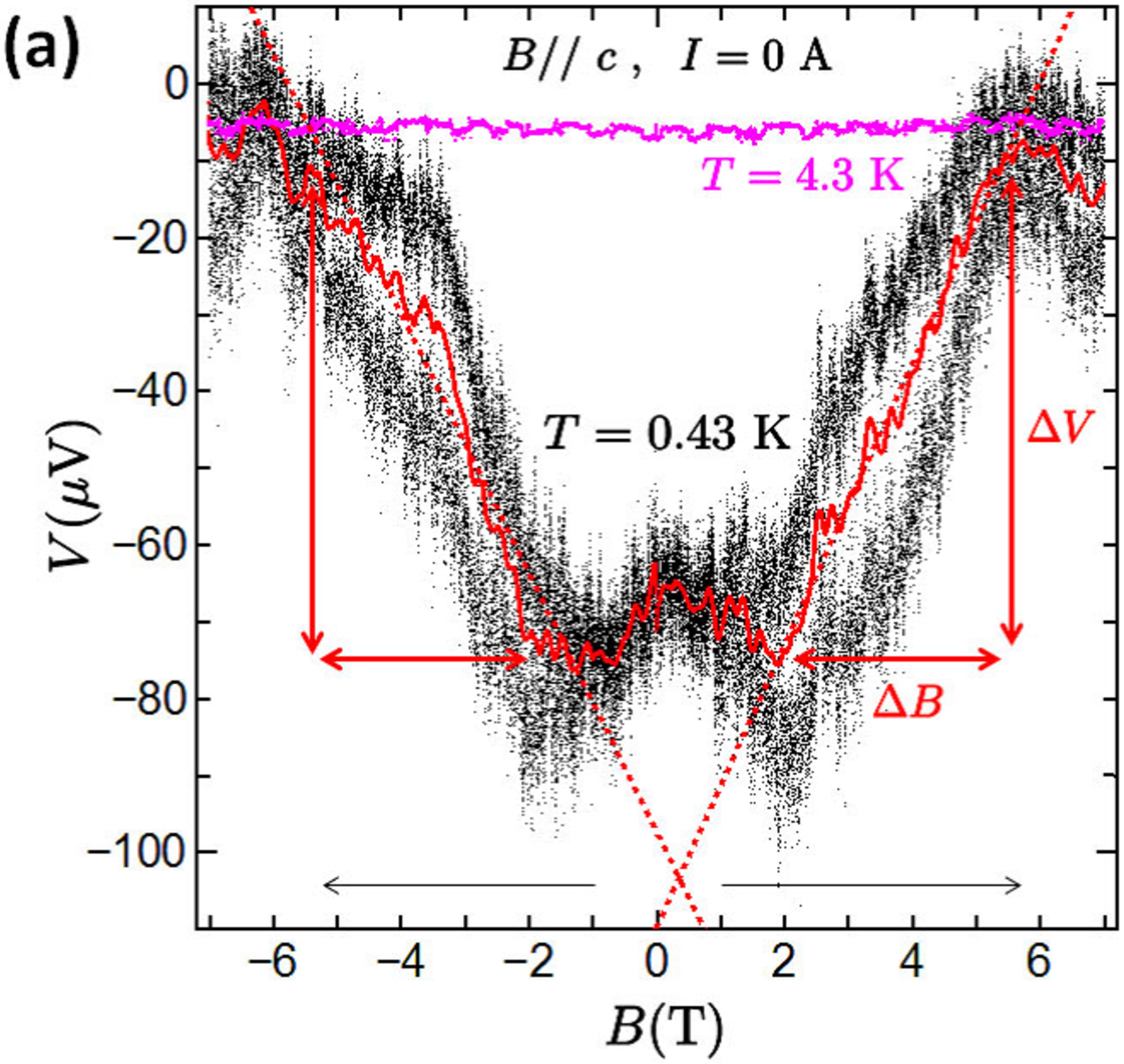}
\includegraphics[width=0.48\linewidth]{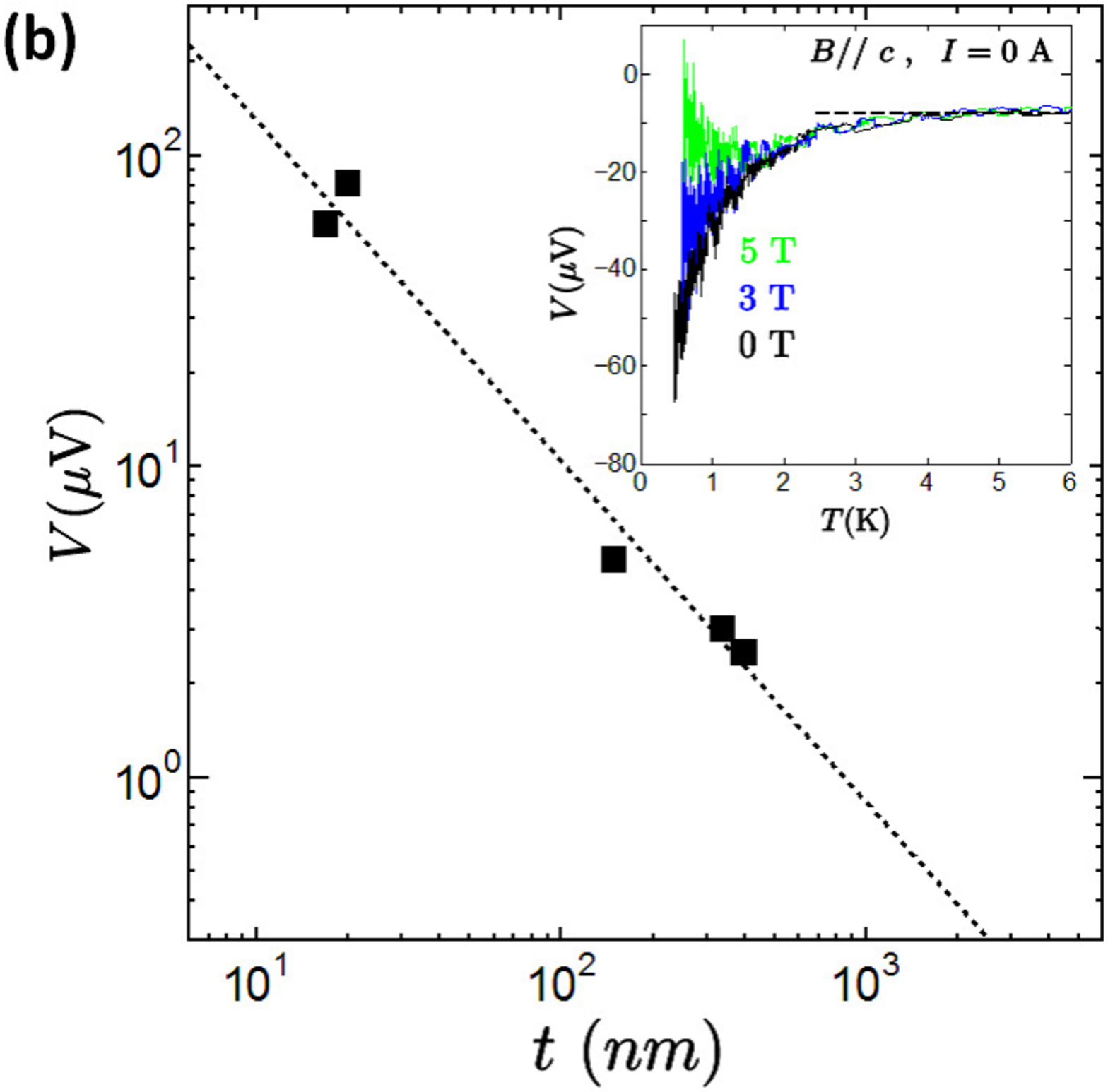}
\includegraphics[width=0.48\linewidth]{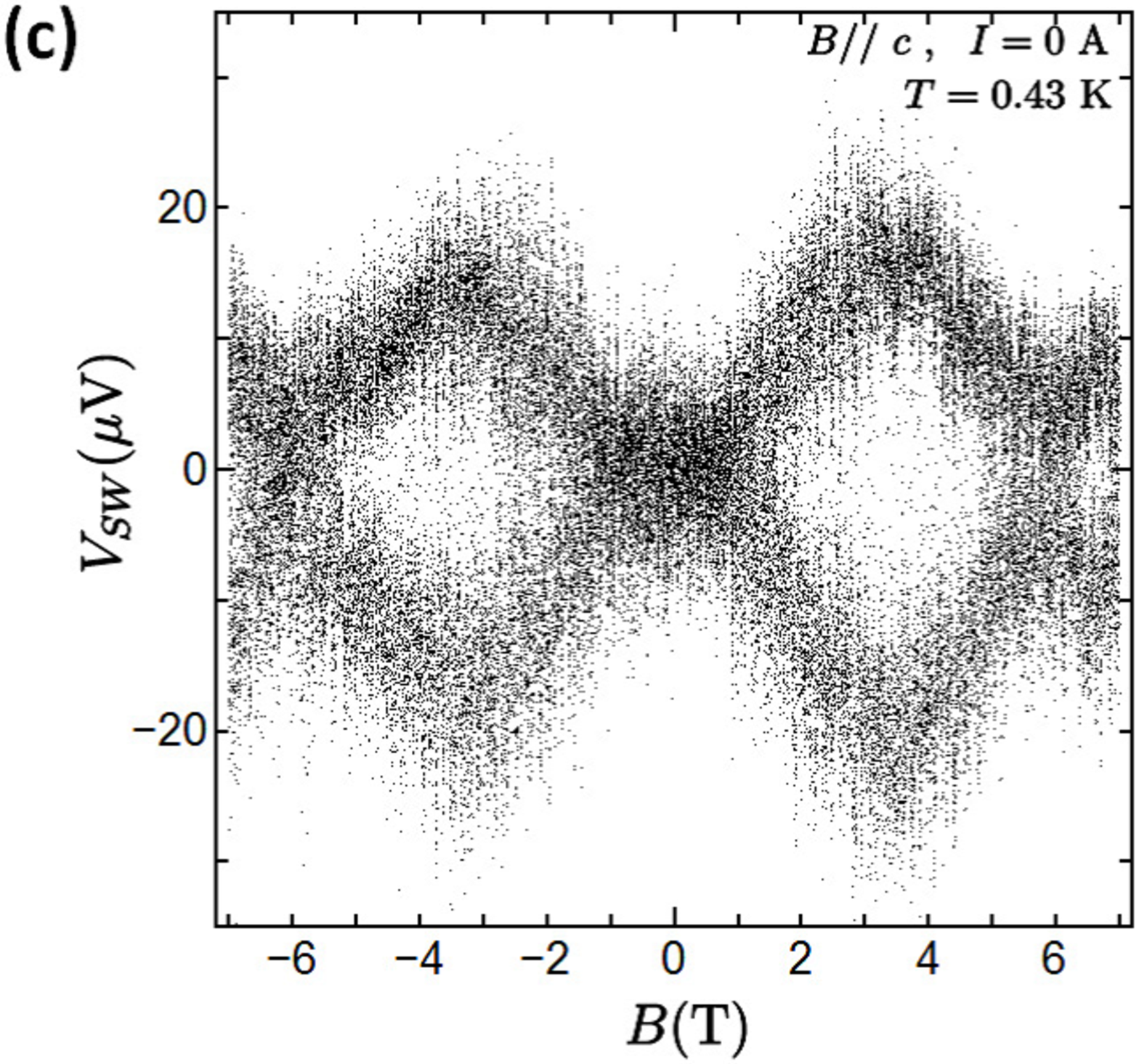}
\includegraphics[width=0.48\linewidth]{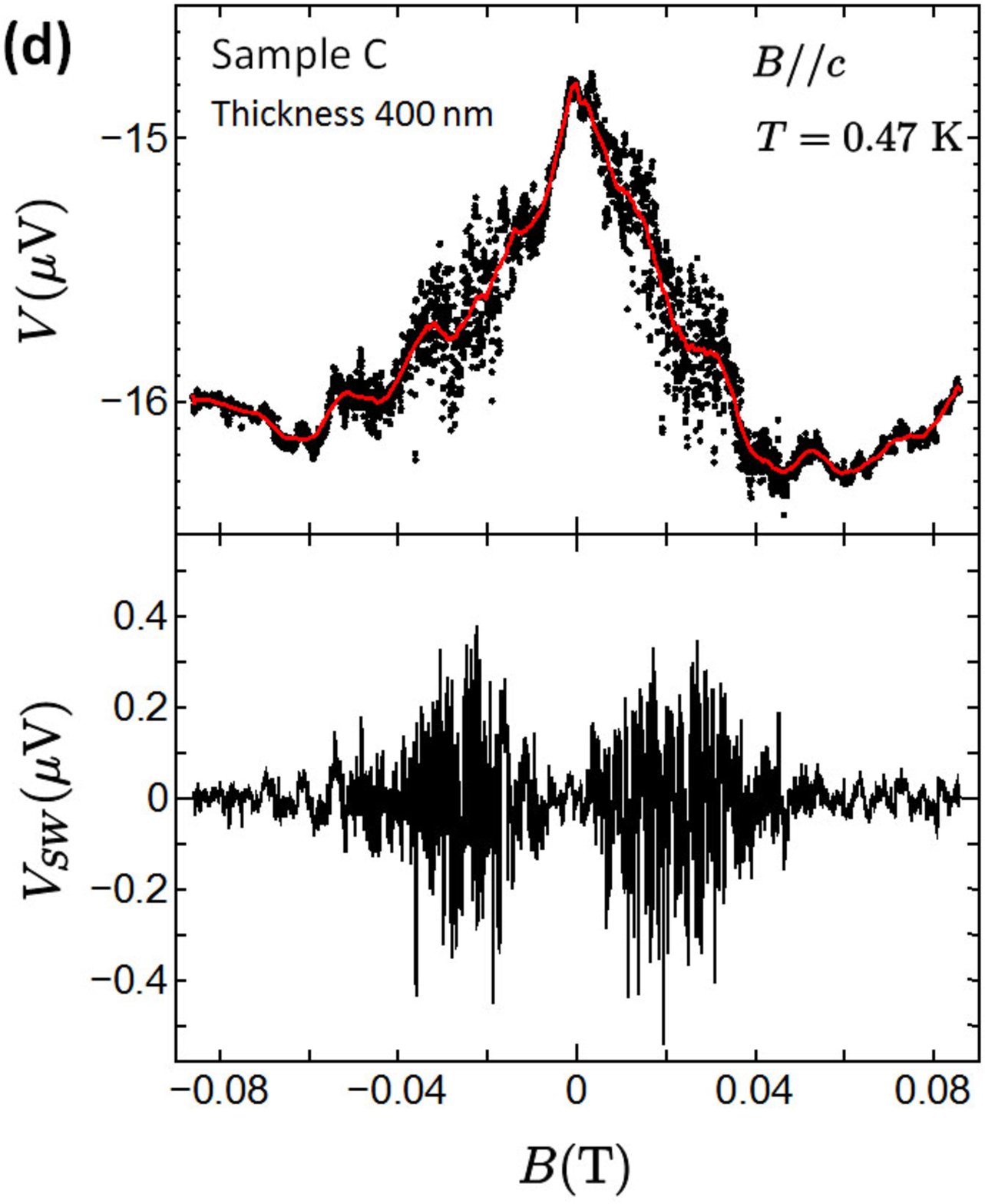}
\caption{
(a)Dependence of the induced voltage on magnetic field at 0.43 and 4.3~K with zero bias current. 
Arrows represent the magnetic sweep direction from zero magnetic field to $\pm~7$~T.
The solid red curve represents the averaged result for the measured data.
(b)Thickness dependence of the induced voltage at lower temperature. 
The dotted line represents the fitting result, which is described well by $V=1/t$. 
The inset shows the temperature dependence of the induced voltage.
(c)The anomalous switching voltage $V_{SW}$ extracted from the $B-V$ curves in Fig~\ref{figure2}(a).
(d)Magnetic field dependence of $V$ and $V_{SW}$ for sample C.
}

\label{figure2}
\end{center}
\end{figure}

\newpage

\begin{figure}[t]
\begin{center}
\includegraphics[width=0.4\linewidth]{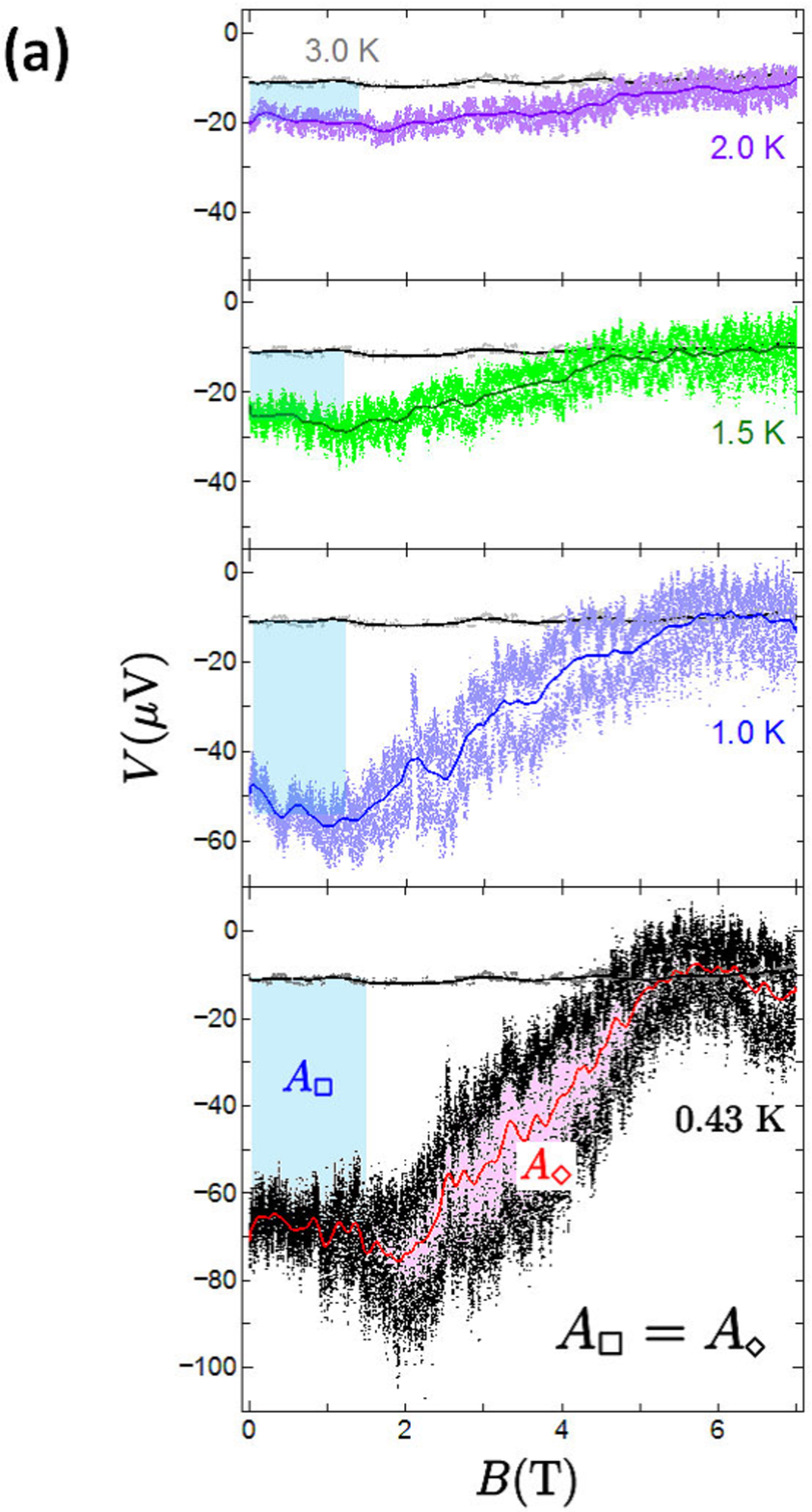}
\hspace{0.3cm}
\includegraphics[width=0.4\linewidth]{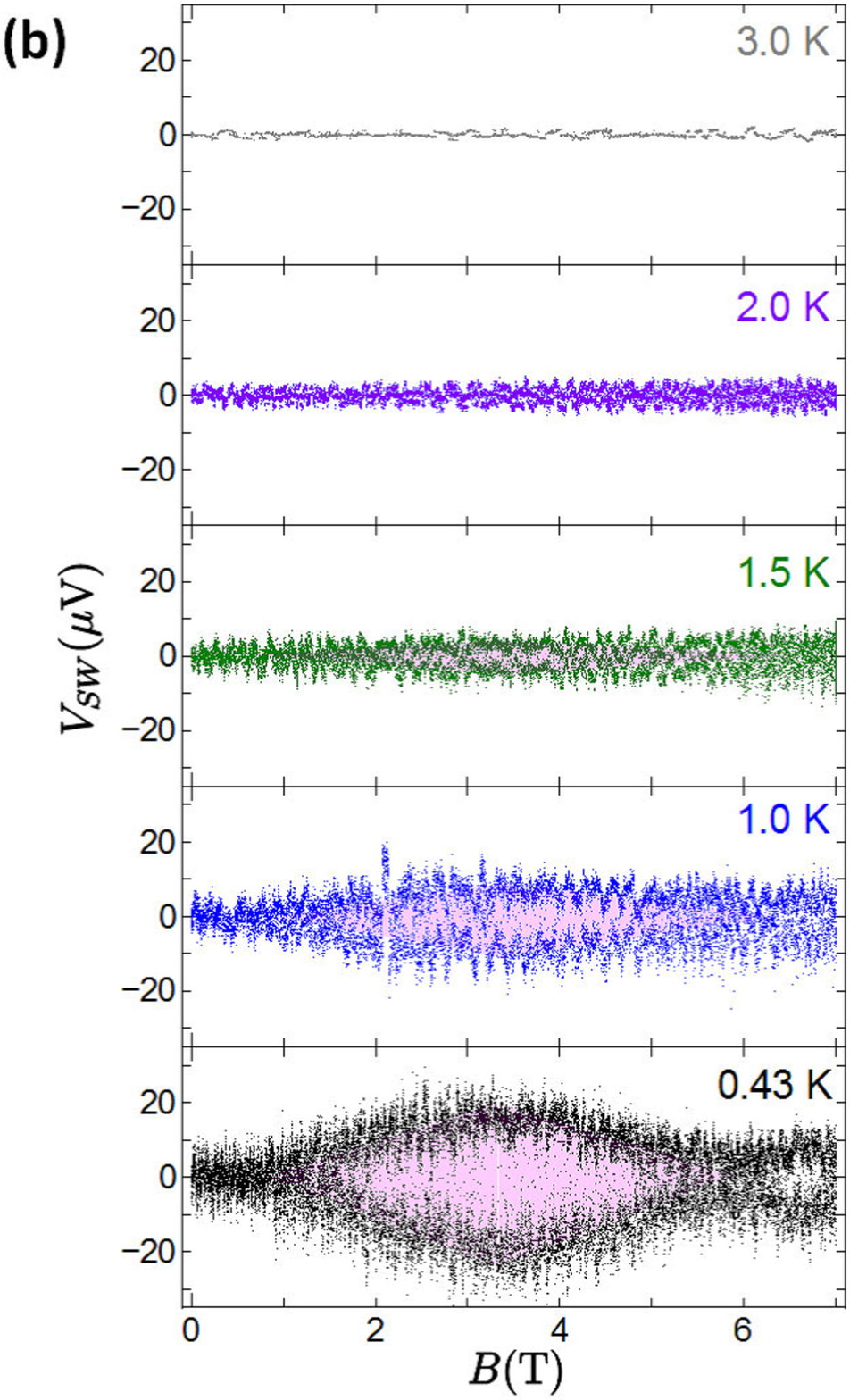}
\includegraphics[width=0.35\linewidth]{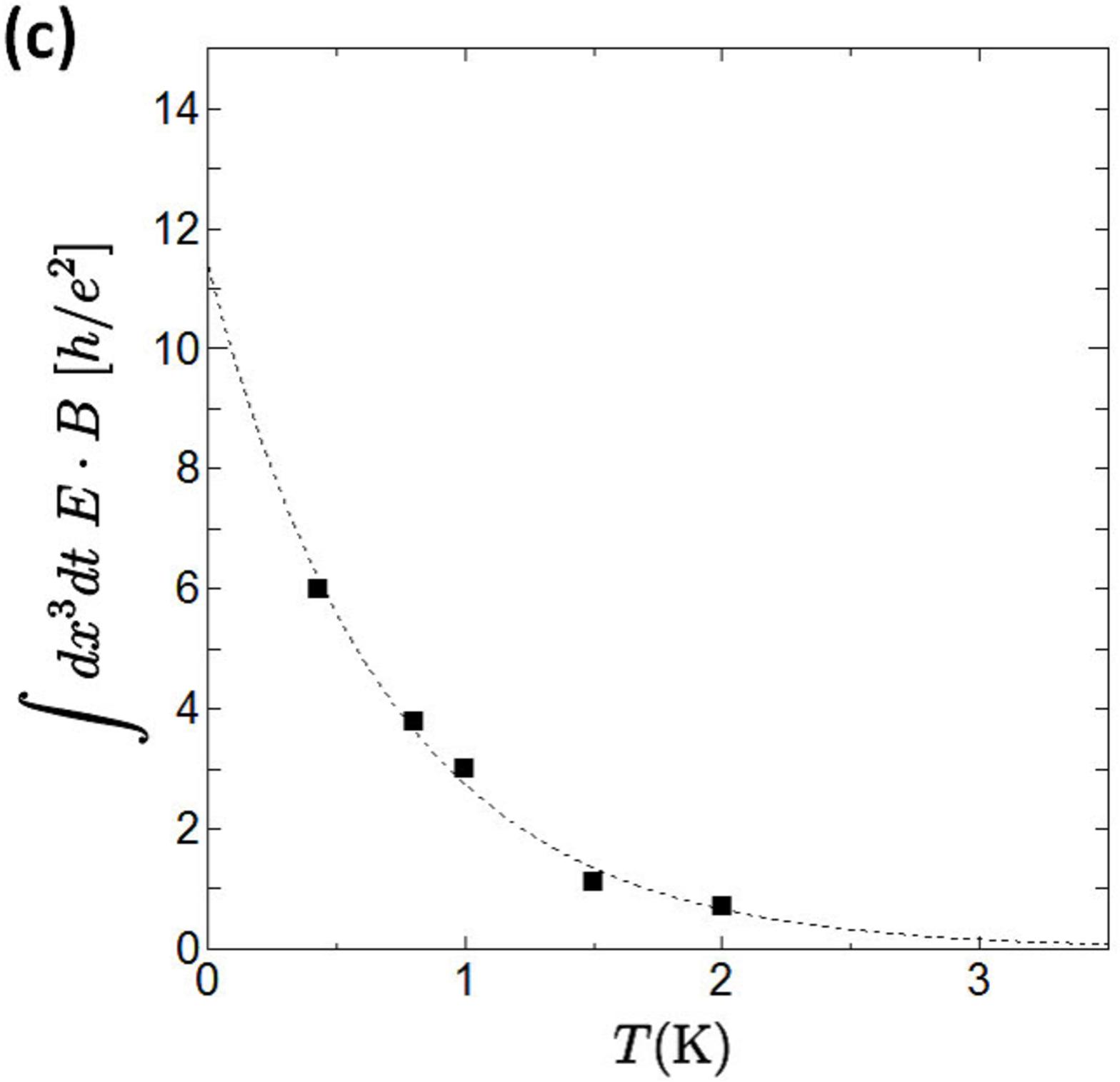}
\hspace{0.3cm}
\label{figure3}
\caption{
(a)Magnetic field dependence of the anomalous voltage at various temperatures with zero bias current from zero magnetic field to 7~T. 
The $E\cdot B$ value in a low magnetic field (the blue region $A_\Box$) is equivalent to that in the red region $A_\Diamond$ caused by voltage switching at temperatures below 2~K.
(b)Temperature dependence of $V_{SW}$.
(c)Temperature dependence of the $E \cdot B$ integral value. The data were fitted by an exponential curve.
}
\label{figure3}
\end{center}
\end{figure}

\end{document}